\documentclass[graybox,vecphys]{svmult}

\usepackage[T1]{fontenc}
\usepackage{mathptmx}       % selects Times Roman as basic font
\usepackage{helvet}         % selects Helvetica as sans-serif font
\usepackage{courier}        % selects Courier as typewriter font
\usepackage{type1cm}        % activate if the above 3 fonts are
\usepackage{makeidx}         % allows index generation
\usepackage{graphicx}        % standard LaTeX graphics tool
\usepackage{multicol}        % used for the two-column index
\usepackage[bottom]{footmisc}% places footnotes at page bottom

\graphicspath{{figures/}}
\usepackage{amsmath}
\usepackage{bbm}
\usepackage{placeins}
\usepackage{float}
\usepackage{graphicx}
\usepackage{hyperref}
\usepackage{physics}
\usepackage{tikz}
\usetikzlibrary{shapes.geometric, quantikz}		%for quantum circuit
\usepackage{adjustbox}			% to adjust quantum circuit size
\usepackage{subcaption}

\usepackage{array}		% for horizontal alignment of tables
\usepackage{multirow}	% for grouping rows

\usepackage{amssymb}
\usepackage{xcolor}
\usepackage{graphicx} 
\usepackage{hyperref}
\usepackage{mathrsfs}		% for \mathscr

\usepackage{url}
\makeindex             % used for the subject index
\DeclareMathAlphabet      {\mathbfit}{OML}{cmm}{b}{it}
\begin{document}

\renewcommand{\vec}[1]{\mathbfit{#1}}
\renewcommand{\hbar}{\mathchar'26\,\mkern-9mu h}

\title*{Hamiltonian engineering with time-ordered evolution for unitary control of electron spins in semiconductor quantum dots}
\titlerunning{Hamiltonian engineering for unitary control of electron spins in quantum dots}

\author{Bohdan Khromets, Zach D. Merino and Jonathan Baugh}

\institute{Bohdan Khromets \at University of Waterloo, N2L 3G1, ON, Canada, \email{bohdan.khromets@uwaterloo.ca}
\and Zach D. Merino \at University of Waterloo, N2L 3G1, ON, Canada, \email{zmerino@uwaterloo.ca}
\and Jonathan Baugh \at University of Waterloo, N2L 3G1, ON, Canada, \email{baugh@uwaterloo.ca}}
\maketitle

\abstract{
	We present a unitary control pulse design method for a scalable quantum computer architecture based on electron spins in lateral quantum dots. 
	We employ simultaneous control of spin interactions and derive the functional forms of spin Hamiltonian parameter pulses for a universal set of 1- and 2-qubit logic gates.
	This includes selective spin rotations with the weak local g-factor variations in the presence of the global oscillating field, and a Control-Phase operation with the simultaneous control of g-factors and exchange couplings. 
	We outline how to generalize the control scheme to multiqubit gate operations and the case of constrained or imperfect control of the Hamiltonian parameters. 
	%	 Our Hamiltonian engineering technique guarantees the time-orderedness of the evolution operator, and the physical system provides an abundance of spin interactions under direct experimental control: electron g-factors, exchange couplings, and global ESR field. This enables us to exactly find the functional forms of spin Hamiltonian parameter pulses for a universal set of 1- and 2-qubit logic gates. This includes the individual control of spins in the global ESR field, and a Control-Phase with the simultaneous control of g-factors and exchange.
	%%%%%%%%%
	\iffalse
	 We demonstrate scalability of the method by designing a 3-qubit Control-SWAP gate numerically with high fidelity, using a few-parameter numerical optimization. 
	Lastly, we provide a pathway towards generalizing the class of Hamiltonians that preserve time-orderedness, and outline how the freedom in choosing the pulse shape can be used to incorporate constraints or  decoupling.
	\fi
	}

\section{Introduction}\label{sec:introduction}

Among the paths towards large-scale quantum computing, spin-based quantum information processing in semiconductor quantum dots is particularly enticing. This is due to large coherence times %\cite{Tyryshkin2011Electronspincoherence}
 in isotopically purified materials like Si or Ge, the possibility of operation at relatively high temperatures \cite{Vandersypen2017Interfacingspinqubits} (1-4 K), and the highly developed MOS/CMOS fabrication technology for semiconductor electronics \cite{Maurand2016CMOSsiliconspin}. The recent demonstrations of fidelity exceeding the fault tolerance threshold for one-\cite{Yoneda2017quantumdotspin} and two-qubit quantum logic gates \cite{Mills2019Shuttlingsinglecharge} boost hope for efficient, universal, and scalable computers in the foreseeable future.

Still, high values of fault-tolerance thresholds and susceptibility to pulse mis-settings and charge noise are significant bottlenecks on the path toward scalability. The development of reliable and precise optimal control algorithms for Hamiltonian engineering is thus an important direction to tackle these problems.
In this work, we establish an analytical pulse design method for unitary evolution, applicable to a specific architecture of semiconductor quantum dot spin qubit processors \cite{Buonacorsi_2019} and compatible with the simultaneous  manipulation of different spin interactions.
%
%This, in particular, includes cases of qubit coupling being comparable or much stronger than the single-electron couplings. 
Certain parts of the proposed method are an extension of quantum circuit physical optimization ideas from \cite{Burkard_1999} to larger functional spaces. 
The analytical nature of our approach  bears similarity to the proposal of Barnes \cite{Barnes_2013} for the exact evolution operator design with the reverse engineering of the Hamiltonian. 
The SMART protocol designed for similar physical systems \cite{Hansen2021Pulseengineeringglobal} follows a different logic, and yet, the freedom to choose the pulse functional profiles offered by our method makes it possible to incorporate findings from other proposals. 
We expect this work to be instrumental in both the control of experimental devices and the numerical engineering of pulses with improved properties  \cite{Riaz2019Optimalcontrolmethods,Yang2019Siliconqubitfidelities} (e.g., noise robustness).

\section{Spin interactions in a scalable quantum dot architecture}\label{sec:theory}
We focus on the realization of a universal, scalable semiconductor spin-qubit architecture \cite{Buonacorsi_2019} based on electron spins in Si/SiO$_2$ lateral quantum dots. It consists of a grid of few-qubit computational nodes, where electron states are controlled locally with voltages on electrodes and {globally} with two perpendicular (static and oscillatory) magnetic fields. Entangled states are distributed between the nodes by means of electron shuttling. The entire rectangular network of nodes undergoes error correction according to the surface code  protocol \cite{Fowler2012SurfacecodesTowards}.

\begin{figure}[h]
	\sidecaption
	\includegraphics[width=0.6\linewidth]{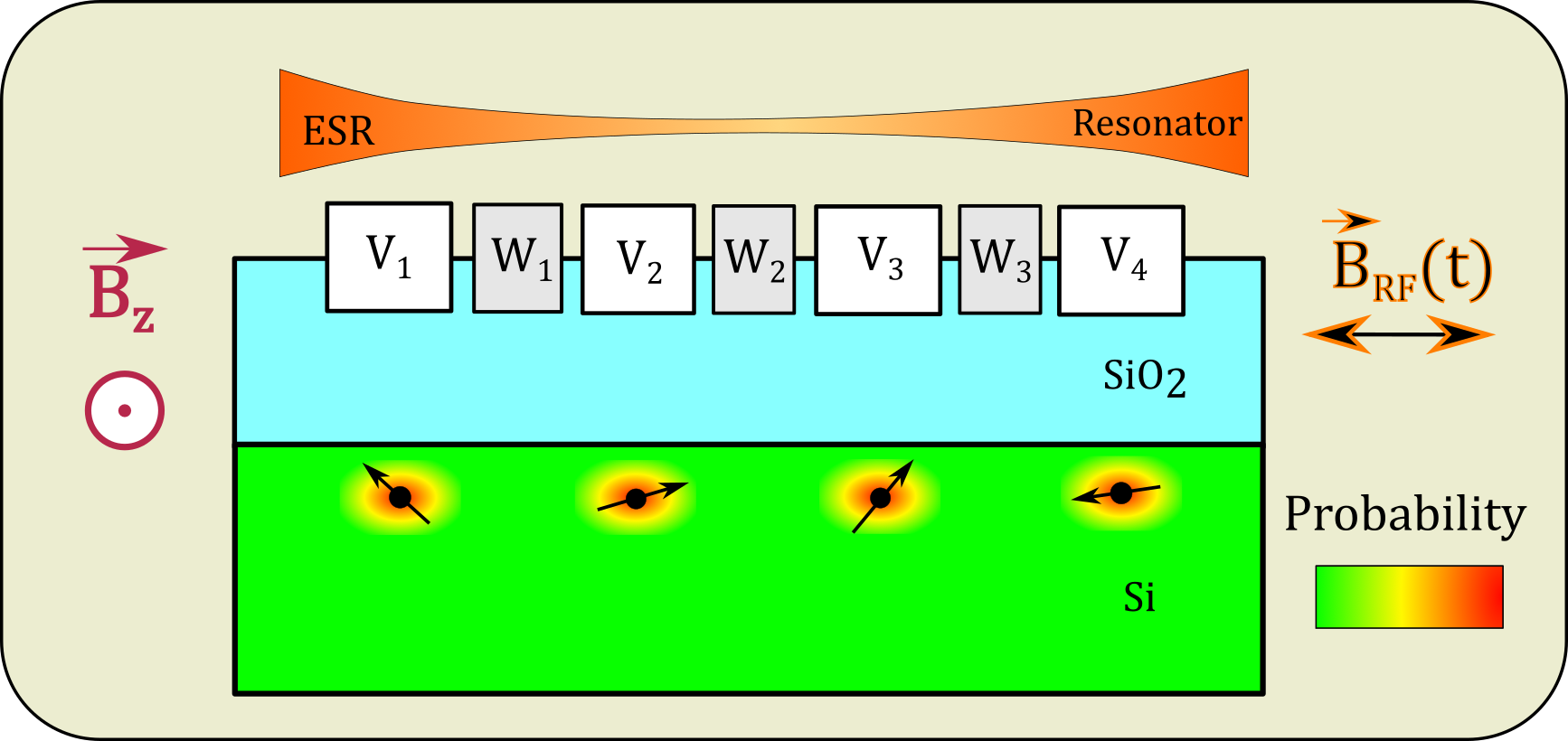}
	\caption[Device schematic]{Schematic of a computational node. Electrons are accumulated in the quantum dots at Si/SiO$_2$ interface and coupled to two (static and oscillatory) magnetic fields. The two-electron interactions and Larmor frequencies are voltage controlled.}
	\label{fig:deviceschematic}
\end{figure}

The schematic of a computational node is presented in figure \ref{fig:deviceschematic}.
A minimal node contains a linear array of four quantum dots formed near the semiconductor (Si) - insulator (SiO$_2$) interface: one for the data qubit (leftmost), two for ancilla qubits (middle), and one for the transfer of electrons between the nodes. Plunger gate voltages $V_i$ accumulate quantum dots and host electrons in them, whereas the tunneling gate voltages $W_i$ control the potential barriers between the dots (and thus the spin coupling strengths).
All spins are identically coupled to two perpendicular global fields: a static Zeeman field $\vec{B}_z$ and an oscillating field $\vec{B}_{\textsc{rf}}(t)$ for electron spin resonance (ESR).

Our goal is to establish a procedure for the coherent control of electron spins in such interacting quantum dot systems. The spin Hamiltonian of a linear array of $N$ spins in the laboratory frame reads:
\begin{multline}
	{H}_\textsc{l}(t)=
	\sum_{j=1}^{N}\underbrace{
		\frac{g_j\left(\vec{V}(t), \vec{W}(t)\right)}{2}\mu_\textsc{b} B_{z} Z_j }_{\text{Zeeman}}	+ 	\sum_{j=1}^{N-1} \underbrace{\frac{J_{j,j+1}\left(\vec{V}(t), \vec{W}(t)\right)}{4}\vec{\sigma}_j \cdot \vec{\sigma}_{j+1}}_{\text{exchange}}
	\\
	+ \underbrace{	\mu_\textsc{b}	B_\textsc{rf}(t) 	\sum_{j=1}^{N}\cos(\int_{0}^{t}\omega_\textsc{rf}(t')\dd{t'}+\phi(t))X_j }_{\text{ ESR with global oscillating field}}. 
	\label{eq:h_lab}
\end{multline}
Here, $X_i,Y_i,Z_i$ denote Pauli operators of the i$^\text{th}$ spin, $\vec{\sigma}_i \cdot \vec{\sigma}_{j}= X_i X_j + Y_i Y_j + Z_i Z_j$, and $\mu_\textsc{b}$ is Bohr magneton. The Hamiltonian terms describe different spin couplings: 

\begin{enumerate}
	\item Voltage-induced $g$-factor deviations introduce slight
	 variability of the qubit Larmor frequencies ${g_j\mu_\textsc{B}B_z}/{2\hbar}$. 
	This is a consequence of a weak but nonzero spin-orbit coupling in Si. 
	% The range of $g$-factor tunability is on the order of $10^{-5}-10^{-4}$.
	\item Direct exchange interaction $J_{i,j}$ between i$^\text{th}$ and j$^\text{th}$ electron arises from Pauli principle and Coulomb repulsion. This coupling is strongly (exponentially) dependent on the tunnel barrier height and electron separation.
	\item Global ESR field is characterized by its time-dependent envelope $B_\textsc{rf}(t)$, angular frequency $\omega_\textsc{rf}(t)$, and phase $\phi(t)$.
\end{enumerate}
We now move to the rotating frame synchronized with the global ESR field. Then, the Hamiltonian~\eqref{eq:h_lab} and evolution operator will read:
\begin{gather}
	R=\exp[\frac{i}{2}\int_{0}^t\omega_\textsc{rf}(t')\dd{t'} \sum_{j=1}^{N} Z_j], \quad U_\textsc{r} = U_\textsc{l} R,\quad H_\textsc{r} = R H_\textsc{l}R^\dagger+i\hbar\dot{R}R^\dagger.
	\label{eq:rot_trans}
\end{gather}
The explicit substitution of the formula for $R$ gives the Hamiltonian (in frequency units) in the rotating frame:
\begin{multline}	\label{eq:rot_frame_hamiltonian}
	\mathscr{H}=\frac{H_\textsc{r}}{\hbar} = \frac{1}{2}\sum_{j=1}^N  \left[ \frac{\mu_\textsc{b}B_{z}}{\hbar}g_j(t)  -{\omega_\textsc{rf}(t)}   \right]  Z_j 
	+ \sum_{j=1}^{N-1} \frac{J_{j,j+1}(t)}{4\hbar}\vec{\sigma}_j \cdot \vec{\sigma}_{j+1}
	\\
	 + \frac{\mu_\textsc{b} B_\textsc{rf}(t)}{2\hbar}\sum_{j=1}^N \left[\cos\phi(t)\; X_j +\sin\phi(t)\; Y_j\right], 
\end{multline}
Note that the rotating-wave approximation has been applied (fast-oscillating terms are ignored). 
Clearly, the evolution under the lab and rotation frame Hamiltonians is equivalent when the operator $R$ performs an integer number of $Z$-rotations during a pulse of length $T$:
$	\int_{0}^{T} \omega_\textsc{rf}(t)\dd{t} = 2\pi n , \ n \in \mathbb{Z}   \ \Rightarrow \ U_\textsc{r}(T) \equiv U_\textsc{l}(T).$

%%%%%%%%%%%%%%%%%%%%%%%%%%%%%%%%%%%%% TO REUSE %%%
%For the pulse design purposes, however, it might not be necessary in case a  Z-rotation by $\gamma\in[0,2\pi)$ needs to be applied on all qubits in addition to the unitary operation $U_\textsc{r}(T)$. Then the condition changes to: $\	\int_{0}^{T} \omega_\textsc{rf}(t)\dd{t} = \gamma\! \mod 2\pi,$ and no extra manipulations with the qubits is needed to achieve this global Z-rotation.

\section{Hamiltonian engineering with the universal control shape} 
Given the Hamiltonian \eqref{eq:rot_frame_hamiltonian}, our goal is to identify functions $g_j(t), J_{j,j+1}(t), $ and $B_\textsc{rf} (t), \omega_\textsc{rf} (t), \phi(t)$ that give deterministic evolution operators consituting a universal set of quantum logic gates. 
%%%%%%%%% 
\iffalse
Due to the restriction on the global ESR control, only some universal 
1-qubit controls are achieved directly by varying control parameters individually.
A signature of the efficient quantum control is also the ability to realize a wide range of multiqubit gates by controlling the Hamiltonian parameters simultaneously. 
\fi
%%%%%
First, it is important to define the initial and final state of the system, in which no qubit evolution in the rotating frame will take place (we will refer to it as idling state).
Keeping the ESR field turned off and making exchange couplings negligibly small by increasing the tunneling barriers before and after each pulse can be trivially done experimentally. 
The only additional requirement to define an idling state is to tune the quantum dots into the regime where the g-factors of all electrons are the same. Then, by choosing the rotating frame appropriately, we set all Z-terms in the Hamiltonian to zero at the beginning and the end of a control pulse, in addition to all other terms:
\begin{equation}\label{eq:idling_requirements}
	\forall i: g_i(\vec{V}_{idle}, \vec{W}_{idle})- g_{idle}=0,   \qquad \omega_\textsc{rf}(0)=\omega_\textsc{rf}(T) = \frac{\mu_\textsc{b}B_{z}}{\hbar} g_{idle}.
\end{equation}
We note that the SMART protocol \cite{Hansen2021Pulseengineeringglobal} implies the same choice of idling state.

\subsection{Spin interactions as basic generators of evolution}
The most basic quantum operations are easiest to realize by manipulating one part of the Hamiltonian \eqref{eq:rot_frame_hamiltonian} at a time while keeping others equal to zero.
The requirements on the profiles of active pulses then become particularly simple:
\begin{enumerate}
	\item Z-rotations of individual qubits by varying their Larmor frequency only:
	\begin{equation}\label{eq:z-rotations}
		\mathrm{ROT}_j(\hat{z}, \theta_i) = e^{-\frac{i}{2} \theta_j Z_j}: \quad \int_0^T \dd{t} \left[ \frac{\mu_\textsc{b}B_{z}}{\hbar}g_j(t)  -\omega_\textsc{rf}(t)\right] = \theta_j \mod 2\pi.
	\end{equation}

	\item $\mathrm{SWAP}^k$ gates on separated qubit pairs by varying their exchange couplings only. The following holds (up to a global phase) due to $\vec{\sigma}_j \cdot \vec{\sigma}_{j+1} = \frac{1}{2}(\mathbbm{1} + \mathrm{SWAP}_{j,j+1}):$
	\begin{equation}\label{eq:swapk}
		\mathrm{SWAP}_{j,j+1}^{k_j} \equiv e^{-\frac{i}{2}\pi k_j \mathrm{SWAP}_{j,j+1}}: \quad \int_0^T \dd{t}  \frac{J_{j,j+1}(t)}{\hbar} = \pi k_j \mod 2\pi.
	\end{equation}

	\item Rotation of all qubits around an axis within x-y plane $\vec{n}= [\cos\varphi, \sin\varphi, 0]$ by the same angle $\theta$ using the global ESR field only:
	\begin{equation}\label{eq:sync_XYrot}
	\prod_{j=1}^{N}\mathrm{ROT}_j(\vec{n}, \theta) = 	\prod_{j=1}^{N} e^{-\frac{i}{2} \theta\vec{n}\cdot \vec{\sigma}_j}	: \quad \int_0^T \dd{t}  \frac{\mu_\textsc{b} B_\textsc{rf}(t)}{\hbar} = \theta \mod 2\pi,\quad \phi(t)\equiv\varphi.
	\end{equation}
\end{enumerate}
Formulas \eqref{eq:z-rotations}, \eqref{eq:swapk}, and \eqref{eq:sync_XYrot} are merely integral constraints on the control functions $g_j(t), J_{j,j+1}(t),B_\textsc{rf} (t), \omega_\textsc{rf} (t)$. That is, one can choose \textit{any functional profiles}  of the control pulse combinations under the integrals as long as they are properly normalized. 
This observation will help us in design arbitrary single-qubit rotations --- one extra building block needed for the  universal control, and in extending the method to more complex quantum operations, including the ones on multiple qubits. 

\subsection{Addressing individual qubits in the global ESR field}
The fundamental feature of the material systems like Si/SiO$_2$ is the weakness of spin-orbit coupling. This makes the range of $g$-factor variations really small for realistic voltage controls. 
For typical Larmor frequencies $\sim 10$ GHz, their tunability range due to $g$-factor deviations is $\sim 0.1-1$ MHz \cite{Laucht_2015}. Thus, qubits cannot be shifted far from resonance to implement selective spin rotations in a traditional way (phenomenon known as frequency crowding \cite{Seedhouse2021Quantumcomputationprotocol}). 
We notice, however, that the control ranges of Larmor and Rabi frequencies $\mu_\textsc{b} B_\textsc{rf}/\hbar$ are comparable, and identify that: 
\begin{svgraybox}
	The combination of global ESR control and weak local $g$-factor variations makes possible the effective local control of individual spins. 
	Specifically, combining the resonant Rabi drive with the \textit{simultaneous} Larmor frequency manipulation to achieve a $2\pi$ rotation at the end of the pulse restores the initial states of the qubits that should be effectively detuned from resonance. 
\end{svgraybox}
Controlling both $X$- and $Z$-components of the Hamiltonian shifts the axis of rotation, and the deterministic rotation by a desired angle can be achieved if the axis of rotation remains fixed during the pulse. This holds when the control pulses are proportional at all times:
\begin{equation}\label{eq:prop_amplitudes}
	 \forall t: \frac{\mu_\textsc{b}B_{z}}{\hbar}g_j(t)  -\omega_\textsc{rf}(t) \propto \frac{\mu_\textsc{b} B_\textsc{rf}(t)}{\hbar}.
\end{equation}
Note that this constraint is not very limiting either: the control pulses can be chosen arbitrarily so long as their \textit{shapes} are the same. 

With these considerations in mind, we now explicitly derive the pulses for a general $\mathrm{ROT}_{+}(\vec{n}, \theta)$ operation on resonant qubits (labeled as "+"), and simultaneously, a $2\pi$ rotation around some shifted axis $\mathrm{ROT}_{-}(2\pi)$ on the non-resonant qubits (labeled as "-"). 
We rewrite the proportionality condition \eqref{eq:prop_amplitudes} by expressing the control parameters in terms of a generalized control shape $S(t)$ and coefficients $A_{\pm}, \text{\ss}$:
\begin{equation}\label{eq:arb_rotations_controls_shape_function}
	\frac{\mu_\textsc{b}B_{z}}{\hbar}
	g_{\pm, j} (t) -  \omega_\textsc{rf} (t) =
	A_{\pm, j}\; \frac{S(t)}{T},  
	%		\qty\Big( g_{\pm, j} (t) -  g_0 )
	\qquad\qquad
	\frac{\mu_\textsc{b}}{\hbar}B_\textsc{rf}(t) = \text{\ss{}}\frac{{S(t)}}{T}.
\end{equation}
For $S(t)$, one can choose any continuous function that starts and ends at zero (to accommodate realistic experimental conditions) with the average value normalized by 1:
%\begin{equation}\label{eq:shape_function}
$
	 \quad S(0)=S(T) =0,  \quad \left\langle S \right\rangle = \frac{1}{T} \int_{0}^T S(t) \dd{t} =1.
$
%\end{equation}
Each single-qubit part of the Hamiltonian then reads:
\begin{gather}\label{eq:hamiltonian_single_qubit_arbitrary_rotation}
	\mathscr{H}_{\pm, j} = \frac{1}{2}\frac{S(t)}{T} 
	\qty\Big[
	 A_{\pm} Z_j + \text{\ss{}} (\cos\phi X_j + \sin\phi Y_j) 
	]
	= \frac{1}{2}\frac{S(t)}{T}\sqrt{A_{\pm}^2 + \text{\ss{}}^2}\; \vec{\sigma}_j \cdot \vec{n}_{\pm, j}, \\
	\vec{n}_{\pm, j} =\left({\text{\ss{}}}\cos\phi, \ {\text{\ss{}}}\sin\phi, \ {A_\pm } \right)^\text{T} /\sqrt{A_{\pm}^2 + \text{\ss{}}^2}, \quad \norm{\vec{n}_{\pm, j}}=1.
\end{gather} 
We align the axis of rotation of resonant qubits with the desired axis: $\vec{n_{+}} = \vec{n} = \left( n_x, n_y, n_z \right)^\text{T}$, and impose requirements on rotation angles of the qubits of two kinds:
\begin{equation}\label{eq:condition_angle_all_arbitrary_rotation}
	\sqrt{A_{+}^2 + \text{\ss{}}^2} = \left|\theta\right|, \qquad \sqrt{A_{-}^2 +  \text{\ss{}}^2} = 2\pi.
\end{equation}
From these two conditions, we obtain the exact expressions for the  amplitudes $A_{\pm},\text{\ss}$ and thus the pulses implementing a $\mathrm{ROT}(\vec{n}, \theta)$ operation on selected qubits:
\begin{equation}\label{eq:pulses_arbitrary_rotation_final}
	\begin{aligned}
		\begin{pmatrix}
		\frac{\mu_\textsc{b}B_{z}}{\hbar}
		g_{+, j} (t) -  \omega_\textsc{rf} (t) \\[5pt]
			\frac{\mu_\textsc{b}B_{z}}{\hbar}
			g_{-, j} (t) -  \omega_\textsc{rf} (t)  \\[5pt]
			\frac{\mu_\textsc{b}B_\textsc{rf}(t)}{\hbar}
		\end{pmatrix}
		=  \frac{S(t)}{T}
		\begin{pmatrix}\displaystyle
			n_{z} \theta
			\\ \displaystyle
			\pm \sqrt{(2\pi)^2 - \theta^2 \left(n_{x}^2 + n_{y}^2\right)}
			\\   \displaystyle
			\left|\theta\right|\sqrt{n_{x}^2 + n_{y}^2}
		\end{pmatrix}& \\
	\phi(t) = \mathrm{atan2}\qty\Big({n_{y}\; \mathrm{sign}\, S(t)
			\; \mathrm{sign}\, \theta},
		{n_{x}\; \mathrm{sign}\, S(t) \; \mathrm{sign} \,\theta}),\qquad&
	\end{aligned}
\end{equation}
with all exchange couplings equal to zero. 
This expression  additionally takes into account the possible sign change of $S(t)$. The ESR phase then becomes a piecewise constant function instead of a constant to keep the ESR field envelope $B_\textsc{rf}(t)$ always positive (becomes proportional to $|S(t)|$ rather than $S(t)$ in formula \eqref{eq:arb_rotations_controls_shape_function} in this case).

\subsection{General shape function formalism for time-ordered evolution}

In the spirit of the control shape constraint \eqref{eq:arb_rotations_controls_shape_function}, 
we generalize the method by making \textbf{\emph{all}} control parameters of an N-qubit Hamiltonian \eqref{eq:rot_frame_hamiltonian} proportional to each other (except the ESR phase which is kept fixed or at most piecewise constant):
\begin{equation}\label{eq:ham_all_proportional}
	\mathscr{H}(t) = \frac{S(t)}{T} \mathscr{H}_0, \quad \mathscr{H}_0 = \frac{1}{2} \sum_{j=1}^N\left[ A_j Z_j + \text{\ss}\left(\cos\phi X_j + \sin\phi Y_j \right)\right] +  \sum_{j=1}^{N-1} C_j \vec{\sigma}_j\cdot \vec{\sigma}_{j+1},
\end{equation}
Here, $\mathscr{H}_0$ is a constant operator with the coefficients defined by formula \eqref{eq:arb_rotations_controls_shape_function} and the expressions for exchange couplings: $J_{j,j+1}(t) = \hbar C_j S(t)/T$. 
The key feature of such a Hamiltonian is that it commutes with itself at all times: 
%\begin{equation}\label{eq:comm_all_times}
$
	\left[\mathscr{H}(t'), \mathscr{H}(t'')\right] \equiv 0 \quad \forall t', t'' \in [0, T].
	$
%\end{equation}
In this case, the evolution operator is time-ordered and is given by an ordinary matrix exponential:
\begin{equation*}
	U(t) = \exp(-i \int_{0}^{t}\! \mathscr{H}(t) \dd t) = \exp(-i \mathscr{H}_0 \int_{0}^t \frac{S(t')}{T}\dd{t'}), \quad U(T)=\exp(-i \mathscr{H}_0).
\end{equation*}
Consequently, complex multi-qubit gates can be engineered by properly choosing the coefficients of $\mathscr{H}_0$ for any shapes $S(t)$ that are properly offset and normalized.
The notable example is the Control-Phase two-qubit gate $\mathrm{CPHASE}(\alpha)$ that we can implement by simultaneously varying exchange and g-factors.
 Burkard and Loss \cite{Burkard_1999} noticed that the $4\times4$ Hamiltonian of type $\mathscr{H}_0 = \frac{1}{2}(A_1 Z_1 + A_2 Z_2) + \frac{C}{4} \vec{\sigma}_1 \cdot \vec{\sigma}_2$ is block diagonal in the computational basis, and thus its  matrix exponential can be found analytically. The coefficients $A_1,A_2,C$ for a $\mathrm{CPHASE}(\alpha)$ gate, same as the constant pulse amplitudes from \cite{Burkard_1999}:
\begin{equation}\label{eq:cphase parameters}
	\begin{aligned}
		&\mathrm{CPHASE}(\alpha): \\ &\scriptstyle
		\begin{pmatrix}
			1 & \cdot & \cdot & \cdot \\[-8pt]
			\cdot & 1 & \cdot & \cdot \\[-8pt]
			\cdot & \cdot & 1 & \cdot \\[-8pt]
			\cdot & \cdot & \cdot & e^{i\alpha}
		\end{pmatrix}
	\end{aligned}  \quad  C = 2\pi - \alpha, \quad 
	A_{1,2} = \begin{cases}
		\frac{1}{2}\left( \alpha \pm \sqrt{\alpha\left(4\pi-\alpha\right)}\right), &  \alpha\in[0,\pi] \\[8pt]
		-\pi + \dfrac{\alpha \pm \sqrt{\alpha\left(4\pi-\alpha\right)}}{2}, &  \alpha\in[\pi, 2\pi),
	\end{cases}
\end{equation}
can now define smooth pulse profiles with shapes $S(t)$.

\begin{figure}
	\sidecaption
	\begin{tikzpicture}
		\node (fig1) at (0,0)
		{\includegraphics[scale=0.45]{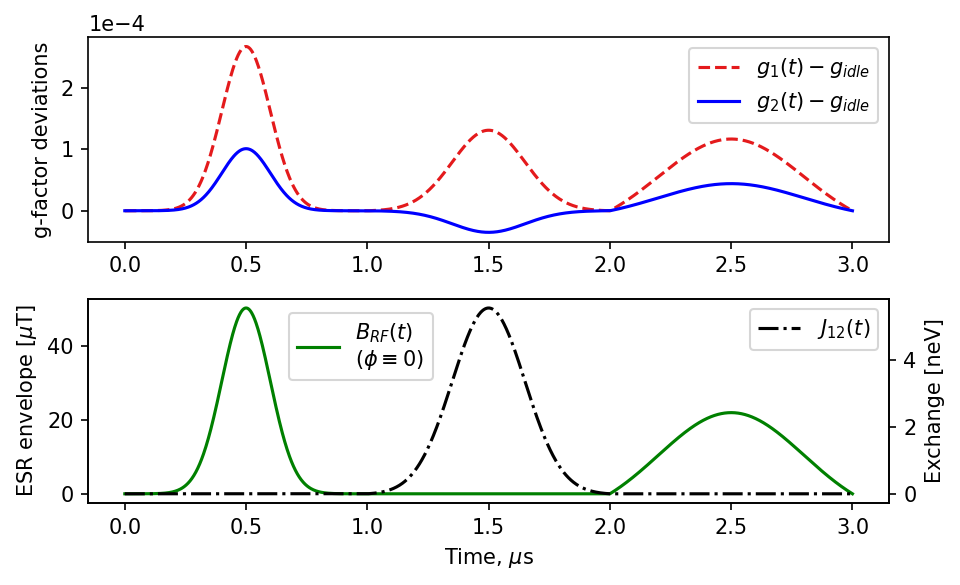}};
		\node (fig2) at (0.05,1.64)
		{		\adjustbox{max width=0.13\linewidth}{
				\colorbox{gray!10!white}{\begin{tikzcd}[row sep=0.1cm]
					\\[0pt]			%looks nicer this way
					& 	\gate{2\pi}	& 	\gate[wires=2]{\mathrm{CZ}}	& 	\gate{2\pi}  & \qw \\
					& \gate{\mathrm{H}}	& 		& \gate{\mathrm{H}}	& \qw \\
			\end{tikzcd} }}
		};  
	\end{tikzpicture}
	\caption[3-pulse sequence for a CNOT gate]{3-pulse sequence for an exact CNOT operation. Each pulse is $1 \mu$s long, and their shapes are Gaussians with $\sigma=0.1,0.15$ and $0.3$, respectively. The system parameters are  $B_z=1$ T, $g_{idle}=2$, and $\omega_\textsc{rf}/2\pi = 27.99248987$ GHz is fixed (given by eq. \eqref{eq:idling_requirements}) }
	\label{fig:cnotsequence}
\end{figure}

To demonstrate our control scheme in action, we find a pulse sequence that realizes a CNOT gate with one control-Z and two Hadamard operations (figure \ref{fig:cnotsequence}). This quantum circuit exemplifies both the selective qubit rotations ($\pi$-rotation around $(101)$ axis for a Hadamard operation, and a $2\pi$-rotation of the non-resonant qubit) and simultaneous control of g-factors and exchange coupling to achieve $\mathrm{CPHASE}(\pi)\equiv \mathrm{CZ}$.
For the control shape functions, we choose normalized Gaussian functions offset vertically to start and end exactly at 0: 
\begin{equation}\label{eq:shifted_gauss}
	S_{\sigma}(\tau) = \left( e^{-\frac{\left(\tau-1/2\right)^2}{2\sigma^2}} - e^{-\frac{1}{8\sigma^2}}\right)/{\int_{0}^{1}\dd{\tau'} \qty\Big[ e^{-\frac{\left(\tau'-1/2\right)^2}{2\sigma^2}} - e^{-\frac{1}{8\sigma^2}} ] } , \quad \tau = \frac{t}{T}\in[0,1].
\end{equation}
We intentionally choose different widths $\sigma$ for each of the 3 pulses in the sequence to emphasize that \textit{any} normalized function for the pulse shape profile will yield an exact unitary operation (i.e., with theoretical fidelity 1).

The quantum logic gate design procedures by choosing shape and coefficients of the Hamiltonian \eqref{eq:rot_frame_hamiltonian} for all gates derived in this section are summarized in Table \ref{tab:parallel_gates}. It also defines two groups of gates that can be run in parallel in quantum circuits, and cases in which the shape functions need not be the same (permitted when the Hamiltonian splits up into spin subspaces).
%input{generalization}
\section{Discussion}
In summary, we devised a pulse design technique for the unitary control of the semiconductor spin qubit processor architecture \cite{Buonacorsi_2019}. The freedom of choosing functional profiles of pulses makes possible to design experimentally relevant smooth pulses that return qubits into the idling state after each operation.
Within the limitations on experimentally controlled spin couplings, a universal gate set of arbitrary spin rotations, and $\mathrm{SWAP}^k$ and control-Phase operations with a theoretical 100\% unitary gate fidelity has been engineered. 
 
The idea of bringing the Hamilonian into the form \eqref{eq:ham_all_proportional} (a product of a scalar function and a constant operator) to make evolution time-ordered, is clearly applicable to systems of any numbers of qubits $N$. Furthermore, it can be directly generalized to other architectures (possibly, with different restrictions on the experimentally controlled parameters) involving, for example, hole spin qubits or the oscillating electric field.  
As the number of  controls available in our architecture $g_j(t), J_{j,j+1}(t)$ grows linearly with $N$ while the Hilbert space size grows as $2^N$, 
the unitary space generated by Hamiltonians of type \eqref{eq:ham_all_proportional} for multiqubit operations becomes progressively more limited. 
Then, by analogy to the proposal for piecewise-constant pulses from \cite{Burkard_1999}, one can design a \textit{sequence} of pulses that generates a desired multiqubit unitary with a certain fidelity by numerically optimizing over the coefficients $A_j, \text{\ss}, C_j$ of each pulse. This procedure will not impact the freedom to choose $S(t)$ for each of the pulses.

Lastly, we note that the design of pulses with improved properties within the proposed control scheme
can be treated as an optimization problem for a function $S(t)$. 
For example, robustness to imperfections of controls due to miscalibrations or noise represented as an ``undesired'' part of the Hamiltonian $\mathscr{H}'(t)$ corresponds to the minimum of the directional derivative of $U(t)$ along  $\mathscr{H}'(t)$ \cite{Haas2019EngineeringEffectiveHamiltonians}. Then, the  optimal $S(t)$ can be found by minimizing  $\norm{\frac{\mathcal{D}U}{\mathcal{D}\mathscr{H'}}}$ \textit{variationally} with respect to  $S(t)$. 
%Together with the restrictions on minimal energy input from ESR field, etc., this problem can be solved by minimizing a corresponding cost function variationally with respect to   or robustness to mis-settings or noise in the Hamiltonian
 Alternatively, one can  use the geometric formalism for dynamically corrected operations \cite{Zeng2019Geometricformalismconstructing,Buterakos2021GeometricalFormalismDynamically} to find pulse shapes with the lowest sensitivity to pulse mis-settings or noise.

\iffalse
Lastly, we ackno
Challenging to calibrate the idling state according to to set all electrons on resonance with each other
and requires either a reliable calibration method, or a control scheme extended to a wider class of functions. 
.
--> \textbf{variational} problem that can be approached analytically. 

Alternatively, approach of Barnes (geometric formalism+)
\fi

\begin{table}[H]
	\caption{Two groups of single- and two qubit gates that can be run in parallel on semiconductor spin quantum processors proposed in \cite{Buonacorsi_2019}}
	\begin{tabular}{m{0.18\linewidth} >{\centering\arraybackslash}p{0.25\linewidth} >{\centering\arraybackslash}p{0.16\linewidth} >{\centering\arraybackslash}p{0.16\linewidth}					m{0.19\linewidth}	}
		\svhline
		\multicolumn{5}{c}{\textbf{Group 1.} Voltage-only driven gates} \\\hline
		Description & Gate layout & Relevant coefficients & Relevant shapes & Requirements \\
		\hline
		Idling qubits  
		& \begin{tikzcd}
			& \qw
		\end{tikzcd}	
		&  \parbox{\linewidth}{
			\begin{gather*}
				A_{-} = 0 
			\end{gather*}
		}
		& 
		&  \multirow{2}{\linewidth}{ 
			\parbox{\linewidth}{
				\vspace{-1.8em}
				No exchange with adjacent qubits:
				\begin{multline*}
					C_{j-1} = \\ C_{j} = 0
				\end{multline*} 	
			}
		}  \\
		\cline{1-4}
		Z rotations by any angles$^a$
		&  \adjustbox{max width=0.9\linewidth}{\begin{tikzcd}
			& 	\gate{\mathrm{ROT(\hat{z},\theta_1)}} & \qw \\[-10pt]
			& 	\gate{\mathrm{ROT(\hat{z},\theta_2)}} & \qw 
		\end{tikzcd} }
		&  $A_{+, j} = \theta_j$
		& $S_{g,j}(t)$ any 
		& \\
		\hline
		Any $\mathrm{SWAP}^k$ gates on separated pairs 
		& \adjustbox{max width=0.9\linewidth}{\begin{tikzcd}[row sep=0.1cm]
			\\[0pt]			%looks nicer this way
			& 	\gate[wires=2]{\mathrm{SWAP}^{k_1}}  & \qw \\[-10pt]
			& 				& \qw \\ 
			& 	\gate[wires=2]{\mathrm{SWAP}^{k_2}}  & \qw \\[-10pt]
			& 				& \qw \\ 
		\end{tikzcd} }
		& \parbox{\linewidth}{ 
			\begin{gather*}
				{C}_{j} = \pi k_j
			\end{gather*} 
		}
		& $S_{{J},j}(t)$ any 
		&  \parbox{\linewidth}{ 
			Unaffected by spin rotations:
			\begin{multline*}
				A_{-,j} = \\ A_{-,j+1} =  0
			\end{multline*} 
		} \\[-0.8em]
	\iffalse %%%%%%%%%%%%%%%%%%%%%%%%%%%%%%%
		\hline
		Hybrid ROT-SWAP gates:  synchronous Z~rotations within each  
		& % \begin{tikzpicture}
			%			\node[scale=0.8, anchor=center] 
			%				{
				\begin{adjustbox}{width=\linewidth}
					\begin{tikzcd}
						\\[-15pt]			%looks nicer this way
						& \gate{\mathrm{ROTZ(\theta_3)}} 
						&	\gate[wires=2]{\mathrm{SWAP}^{k_3}}  & \qw \\
						& \gate{\mathrm{ROTZ(\theta_3)}}	&		& \qw \\ 
					\end{tikzcd}
				\end{adjustbox}
				%};
			%\end{tikzpicture} 
			& \parbox{\linewidth}{
				\begin{gather*}
					A_{+, j} = \\
					A_{+, j+1} = \\ 
					\theta_j 
				\end{gather*}
				\begin{gather*}
					{C}_{j} = \pi k_j
				\end{gather*}
			}
			& \parbox{\linewidth}{
				\vspace{-1.5em}
				\begin{gather*}
					S_{{g},j}(t) = \\ S_{{g},j+1}(t)
				\end{gather*} 
				\\[5pt]
				$S_{J,j}(t)$ any
			} 
			& \parbox{\linewidth}{ 
				\vspace{-1.5em}
				No exchange with the qubits adjacent to the pair:
				\setlength{\abovedisplayskip}{0pt} \setlength{\abovedisplayshortskip}{0pt}
				\begin{multline*}
					C_{j-1} = \\ C_{j+1} =   0
				\end{multline*} 
			}  \\
		\fi %%%%%%%%%%%%%%%%%%%%%%%%%%%%%%%%%%%%%%%%%%%%%%%%%%%%%%%%%%
			\hline
			Any $\mathrm{CPHASE}(\alpha_j)$ gates on separated pairs
			 	& % \begin{tikzpicture}
				%			\node[scale=0.8, anchor=center] 
				%				{
					\adjustbox{max width=0.9\linewidth}{
						\begin{tikzcd}[row sep=0.1cm]
							\\[0pt]			%looks nicer this way
							& 	\gate[wires=2]{\mathrm{CPHASE}(\alpha_1)}  & \qw \\[-10pt]
							& 				& \qw \\ 
							& 	\gate[wires=2]{\mathrm{CPHASE}(\alpha_2)}  & \qw \\[-10pt]
							& 				& \qw \\ 
					\end{tikzcd} }
					%};
				%\end{tikzpicture} 
				& 
				\parbox{\linewidth}{
					\vspace{4pt}
					$ \alpha_j\in[0,\pi]:$
					\begin{gather*}
							A_{+, j/j+1} = \\ 
						\tfrac{\alpha_j \pm \sqrt{\alpha_j (4\pi - \alpha_j )  }}{2}
						\\[0.7em]
							{C}_{j} = 2\pi - \alpha_j
					\end{gather*}
				}
				& \parbox{\linewidth}{
					\begin{gather*}
						S_{{g},j}(t) = \\ S_{{g},j+1}(t)  = \\ S_{J,j}(t)
					\end{gather*} 
					\\[5pt]
				} 
				& \parbox{\linewidth}{ 
					\vspace{-1.5em}
					No exchange with the spins adjacent to the active pairs:
					\setlength{\abovedisplayskip}{0pt} \setlength{\abovedisplayshortskip}{0pt}
					\begin{multline*}
						C_{j-1} = \\ C_{j+1} =   0
					\end{multline*} 
				}  \\[-0.5em]
				\hline
				\multicolumn{2}{c}{Global parameters} & \multicolumn{3}{c}{\ss{} = 0} \\[1em]
			\end{tabular}
		\vspace{1em}
			\begin{tabular}{m{0.18\linewidth}  >{\centering\arraybackslash}p{0.25\linewidth} >{\centering\arraybackslash}p{0.16\linewidth} >{\centering\arraybackslash}p{0.16\linewidth}					m{0.19\linewidth}	}
				\svhline
				\multicolumn{5}{c}{\textbf{Group 2.} Voltage and ESR-driven gates} \\
				\hline
				Description & Gate layout & Relevant coefficients & Relevant shapes & Requirements \\
				\hline
				Synchronous  resonant qubit rotations$^a$  $\mathrm{ROT}(\vec{n}, \theta)$ 
				& 		\adjustbox{max width=0.9\linewidth}{\begin{tikzcd}
					& 	\gate{\mathrm{ROT}(\vec{n}, \theta)} & \qw \\[-10pt]
					& 	\gate{\mathrm{ROT}(\vec{n}, \theta)} & \qw 
				\end{tikzcd} }
				&  \parbox{\linewidth}{
					\begin{multline*}
						A_{+} =  n_z \theta
					\end{multline*}
				}
				& \multirow{2}{\linewidth}{
					\centering
					\vspace{-4em}
					$S_{g}(t) = S_\textsc{b}(t)$
				}
				&  
				\vspace{-2em}\multirow{2}{\linewidth}{
					No exchange with adjacent qubits:
					\begin{multline*}
						C_{j-1} = \\ C_{j} =   0
					\end{multline*} 
				} 
				\\
				\cline{1-3} 
				Synchronous nonresonant  $2\pi$ rotations $^a$
				& \begin{tikzcd}
					& 	\gate{2\pi} & \qw 
				\end{tikzcd} 
				& \parbox{\linewidth}{\vspace{-1em}
					\begin{multline*}
						A_{-} = \\
						\pm \left[4\pi^2 - \theta^2 \times \right. \\
						\left. \left(n_{x}^2 + n_{y}^2\right)\right]^{{1}/{2}}
					\end{multline*}
				}
				&  &  \\[-0.5em]
				\hline
				\iffalse
				ROT-SWAP gates on nonresonant qubits acting as ``pure'' $\mathrm{SWAP}^k$ gates
				&  \begin{adjustbox}{width=\linewidth}
					\begin{tikzcd}
						\\[-15pt]			%looks nicer this way
						&	\gate[wires=2]{\mathrm{SWAP}^{k_1}} 
						& \gate{2\pi} & \qw \\
						& 	&	\gate{2\pi}	& \qw \\ 
					\end{tikzcd}
				\end{adjustbox}
				& \parbox{\linewidth}{
					\centering
					$A_{-}$ same as above
					\begin{gather*}
						{C}_{j} = \pi k_j
					\end{gather*} 
				} 
				& \multirow{2}{\linewidth}{
					\centering
					$S_{g}(t) = S_\textsc{b}(t)$  \\[10pt]
					$S_{J,j }(t)$ any
					
				}
				& \multirow{2}{\linewidth}{
					\parbox{\linewidth}{
						\vspace{-3em}
						No exchange with the qubits adjacent to the pair:
						\begin{multline*}
							C_{j-1} = \\ C_{j+1} =   0
					\end{multline*} }
				} 
				\\
				\cline{1-3}
				Hybrid ROT-SWAP gates on resonant qubits 
				&  \begin{adjustbox}{width=\linewidth}
					\begin{tikzcd}
						\\[-10pt]			%looks nicer this way
						& \gate{\mathrm{ROT}(\vec{n}, \theta)} 
						&	\gate[wires=2]{\mathrm{SWAP}^{k_2}}  & \qw \\
						& \gate{\mathrm{ROT}(\vec{n}, \theta)}	&		& \qw \\ 
					\end{tikzcd}
				\end{adjustbox}  
				&  \parbox[c]{\linewidth}{
					\begin{gather*}
						A_{+} = n_z \theta \\
						C_{j} = \pi k_j
					\end{gather*}
				}
				&  &  \\
				\hline
				\fi
				\multicolumn{2}{c}{Global parameters} 
				& \multicolumn{3}{c}{
					\parbox{0.3\linewidth}{\vspace{-0.8em}
						\begin{align*}
							\text{\ss{} }&= \abs{\theta} \sqrt{n_x^2 + n_y^2} \\
							\phi(t) = \mathrm{atan2}&\left(
							{n_{y}\; \mathrm{sign}\, \theta\; \mathrm{sign}\, S(t) } ,  \right.
							\\
							&\quad \left. {n_{x}\; \mathrm{sign}\, \theta\; \mathrm{sign} \, S(t) }		\right)
						\end{align*}
					}
				}\\[-0.5em]
				\hline
			\end{tabular}
			\label{tab:parallel_gates}
			
			\footnotesize $^a$ due to the fact that $\vec{{n}}\cdot (\vec{\sigma}_1 + \vec{\sigma}_2)$ commutes with $\vec{\sigma}_1 \vec{\sigma}_2$ for any $\vec{n}$, a $\mathrm{SWAP}^k$ gate can run \textit{simultaneously} with any pair of synchronously rotating qubits (i.e., same axis and angle of rotation).
		\end{table}

\iffalse %%%%%%%%%%%%%%%%%%%%%% 

"functional realizations of quantum information processing"
\fi

\bibliography{silicon}

\end{document}